# Elevating Information System Performance: A Deep Dive into Quality Metrics

Dana A Abdullah [1,2], Hewir A. Khidir [7], Ismail Y. Maolood[1], Aso K. Ameen [2,1,] Dana Rasul Hamad [3], Hakem Saed Beitolahi[3] , Abdulhady Abas Abdullah[4] ,Tarik Ahmed Rashid[5] ,Mohammed Y. Shakor[6]

[1] Department of Information and Communication Technology Center (ICTC)- System Information, Ministry of Higher Education and Scientific Research, Kurdistan Region – F.R. Iraq
[2]Department of Computer Science, College of Science, Knowledge University, Erbil 44001, F.R. Iraq
[3] Computer Science Department, Faculty of Science, Soran University, Soran, Erbil, KRG, F.R. Iraq
[4]Artificial Intelligence and Innovation Centre, University of Kurdistan Howler Erbil, Kurdistan Region, F.R. Iraq
[5] Computer Science and Engineering Department, University of Kurdistan Howler, Erbil, Kurdistan Region, F.R. Iraq
[6]Department of Computer, College of Science, University of Sulaimani, Sulaymaniyah 46001,F.R. Iraq
[7] Department of Statistics, College of Administration and Economics, Salahaddin University-Erbil, Kurdistan Region - F.R. Iraq.

Complete authors' information:
1st Author's organizational email: dana.ali@mhe-krg.org
1st Author's ORCID link: https://orcid.org/0009-0009-7610-3157
1st Author's Google Scholar Citation link: https://scholar.google.com/citations?hl=en&user=w60L0LgAAAAJ

2st Author's organizational email: hewir.khidir@su.edu.krd
2st Author's ORCID link: https://orcid.org/0009-0003-6376-1093
2st Author's Google Scholar Citation link: https://scholar.google.com/citations?hl=en&user=1iA66C8AAAAJ

3st Author's organizational email: Ismail.maulood@mhe-krg.org
3st Author's ORCID link: https://orcid.org/0000-0003-1683-1493
3st Author's Google Scholar Citation link: https://scholar.google.com/citations?hl=en&user=tdB8KHQAAAAJ

4st Author's organizational email: aso.khaleel@mhe-krg.org
4st Author's ORCID link: https://orcid.org/0000-0002-7037-7546
4st Author's Google Scholar Citation link: https://scholar.google.com/citations?hl=en&user=3tL2kesAAAAJ

5th Author's organizational email: dana.hamad@soran.edu.iq
5th Author's ORCID link: https:// https://orcid.org/0000-0002-5230-2225
5th Author's Google Scholar Citation link: https://scholar.google.com/citations?hl=en&user=NJ1ulxIAAAAJ

6th Author's organizational email: Hakem.beitollahi@soran.edu.iq
6th Author's ORCID link: https://orcid.org/0000-0002-8420-6545
6th Author's Google Scholar Citation link: https://scholar.google.com/citations?hl=en&user=SMnMHbQAAAAJ

7th Author's organizational email: abdulhady.abas@ukh.edu.krd
7th Author's ORCID link: https://orcid.org/0009-0007-5508-9371
7th Author's Google Scholar Citation link: https://scholar.google.com/citations?user=fxIiXlMAAAAJ&hl=en&oi=ao

8th Author's organizational email: tarik.ahmed@ukh.edu.krd
8th Author's ORCID link: https://orcid.org/0000-0002-8661-258X
8th Author's Google Scholar Citation link: https://scholar.google.com/citations?user=PIPtepUAAAAJ&hl=en

9th Author's organizational email: mohammed.yousif@garmian.edu.krd
9th Author's ORCID link: https://orcid.org/0000-0003-2203-0393

**ABSTRACT**

**In today's digital age, information systems (IS) are indispensable tools for organizations of all sizes. The quality of these systems, encompassing system, information, and service dimensions, significantly impacts organizational performance. This study investigates the intricate relationships between these three quality dimensions and their collective influence on key performance indicators such as customer satisfaction and operational efficiency. By conducting a comparative analysis of various quality metrics, we aim to identify the most effective indicators for assessing IS quality. Our research contributes to the field by providing actionable insights for researchers or practitioners to develop the implementation, evaluation and design of information systems. Also, a quantitative study employing a structured questionnaire survey was conducted to achieve primary data from respondents across various sectors. Statistical analysis, including Cronbach's Alpha (0.953) and factor analysis (KMO = 0.965, Bartlett's Test p < 0.000), revealed strong interdependencies among System Quality (SQ), Information Quality (IQ), and Service Quality (SerQ). The results demonstrate that high SQ leads to improved IQ, which in turn contributes to enhanced SerQ and user satisfaction. While all three qualities are crucial, SerQ emerges as the most relevant indicator of overall system performance due to its broader representation of quality dimensions.**

## I. INTRODUCTION

Information systems have emerged as essential assets for organizations to undertake optimal operations, make appropriate decisions, and gain competitive advantage (Ameen, Kadir, et al., 2024). In the present-day world, information systems play a central role in business and organizational processes of managing work, data and creating new products and services. Nevertheless, as information systems become larger and more sophisticated, they present challenges like poor data quality, system failure and user dissatisfaction (Ameen, Kadir, et al., 2024). To solve these problems, and thus guarantee the success of information systems, one needs to focus on the quality characteristics of these systems. When quality becomes the goal, organizations can optimize system performance, increase the satisfaction of users and consequently, reach strategic objectives.

There are several factors that have been proposed to affect the performance of system of Information system among them are: (SQ), (IQ) and (SerQ) (San et al., 2020). SQ can be defined as the technological attributes of the system with emphasize on the efficiency of the operation, its efficacy as well as the usability of the system. As for the definition of a high SQ system, one that is, quick, reliable, and convenient. IQ relates information quality to the nature, precision, comprehensiveness, and up-to-date character of the data being processed by the system (Ameen, Maolood, et al., 2024). The present study will use high-quality information as a way of making efficient decisions and solving existing problems. The text here can be clicked/tapped to enter the information required. This change lays down emphasis on the fact that the key to sound decisions, as well as stellar performance across organizational functions, lies in the type of information provided. SerQ based on the SERVQUAL model encompasses user services regarding availability, courtesy, and comprehension. The SERVQUAL model, employed by (Nde et al., 2010),continues to be an essential tactic for measuring service quality across different sectors. High service quality brings the high level of consumers' satisfaction, trust, and further cooperation.

The purpose of this study is to explore the nature of interactions among the identified quality dimensions as well as their effect on organizational performance. Recognizing the relationships between SQ, IQ and SerQ allow strategizing the organizational improvements in question, system dependability, as well as data relevance and user satisfaction. This research purposes to determine the factors that affect each of the quality dimensions, analyze the correlation between SQ, IQ, and SerQ, and analyses the effect of the quality dimensions on organizational performance variables, including customer satisfaction, operations, and financial performance as well as to propose an approach for evaluating and enhancing IS quality. In answering these research questions this study is able to add to the knowledge base in the field of information systems and offers best practices to organizations that wish to improve the quality of their systems.

The research design used was quantitative, with the target population of 236 respondents from the different sectors. A designed questionnaire was developed to collect data on the three quality dimensions: SQ, IQ, and SerQ. The questionnaire was conducted through online survey among a diverse respondent group which consisted of IT specialists, business analysts, and end-users. Cronbach's Alpha coefficient of (0.953) and factor analysis with (KMO value of = 0.965 and Bartlett's Test value of $p < 0.000$) indicated the reliability and validity of the data. Also, to estimate the local consistency of the scales Cronbach's Alpha was utilized and to establish the factors underlying the constructs taking advantage of factor analysis.

The objective of this study is to contribute to the literature on quality metrics in the context of information systems through findings from this paper. For academic scholars and practitioners, this research offers valuable understanding by conceptualizing a holistic assessment framework. Results: The results provide suggestions for enhancing system reliability, information trust and user satisfaction in organizational setting. The answers to these questions can also be great indicators to guide decisions in the development, implementation, and maintenance of systems resulting in benefits for other companies, which are: better organizational performance with better information system performance. The paper contributes to the literature along several dimensions.

- Comprehensive Framework: It offers a systematic review index for evaluating the quality of information system by including SQ, IQ, and SerQ.
- Interrelationships: It investigates the interrelationships between SQ, IQ, and SerQ, highlighting their mutual influence on system performance.
- Relative Importance: This allows the comparing of the relative emphasis given to such measures as SQ, IQ, and SerQ that define SerQ as one of the most valuable assessments of the total system performance.
- Practical Implications: It is beneficial for all those organizations that would like to attain system reliability, accurate information flows, and high levels of services in a short time.
- Theoretical Contributions: It advances the theoretical knowledge of information systems quality by offering support for the connections between components of SQ, IQ, and SerQ.

The remainder of the paper is as follows: Section II emphasis on the literature review. Section III brings out the research methodology Section IV explicates on statistical data and analysis. In Section V, the authors describe simulation outcomes. The last section of this paper is section VI which afford an overall discussion while section VII offers the conclusion of the paper.

## II. LITERATURE REVIEW

This section reviews prior studies on three quadrants of SQ, IQ, and SerQ to determine how these factors provide a basis for evaluating the quality and accuracy of information systems. Measurement quality metrics serve as a basis in the understanding of status of information systems and their relevance to organizational objectives (van de Ven et al., 2023). Prior works have also pointed to these metrics as being significant across different settings such as customer outcomes, operational efficiency and organizational performance. (Cvjetković et al., 2021). Also, (Al-Kofahi et al., n.d.) introduced the IS success model, emphasizes the importance of SQ as a factor that impacts on the success of information systems. Research has extended this model in various ways focusing on features like system dependability, performance time, and usability of the system. (Oktrivina et al., 2021a). Moreover,

(Camilleri, 2024) and (Alsyouf et al., 2023) pointing out that reliability, usability, security, and performance efficiency are referred to as the key elements of system quality for information systems. From these studies it is clear that SQ is an important enabler for the effectiveness and efficiency of IS. (Ferreira et al., 2020) went further and determined that both functionality and usability effected to satisfy the users and understood organizational usefulness of the system. This is in line with the current study's aim of assessing SQ as a positive factor in information system success. (Almutairi & Aljohani, 2024) further enhanced the discussion of reliability and performance criteria of system quality that is further defined by availability of the system, the reliability rate as well as the speed of implementation of solutions. Such points suggest that SQ is a rather complex concept that can influence both the utilization of information systems by the user and the organizational performance.

IQ involves the end product that is generated by the system which can be measured with regard to aspects such as accuracy, completeness, relevancy and time sensitivity. A wide-range framework has been provided by (Stvilia et al., 2007) to evaluate IQ, identifying these dimensions as crucial for decision-making processes. For example, (Garrido-Moreno et al., 2024) emphasized the issue of time and detail in generating quality information to complement the managerial decision-making process and organizational performance. Lack of information quality is likely cause for mistakes, ineffective processes, and in general negative organizational consequences. This work adopted accuracy, completeness, relevance, and timeliness as sub-elements of IQ in determining the overall system quality. (Kalankesh et al., 2020) established that informational relevance and comprehensiveness have high quality of system performance and positive effectiveness to persuade the users. Because of these findings, therefore, the role of IQ as one of the system quality attributes receives further support. Thus, the presented study combines dimensions like accuracy, completeness, relevance, and time relevance in order to provide a more extensive the quality of information evaluation, provided by information systems.

Endpoints of SerQ relate to the communication for the service users and the service providers in the area of perceived attentiveness, frequency of communication, friendliness amongst others. The SERVQUAL model introduced in this paper still holds the importance as a significant framework to measure the service quality in different industries. (Johnson et al., 2018) has highlighted the role of Responsiveness and Assurance in forming the trust and satisfaction of the users. Unfortunately, a poor SerQ can lead to customer complaints, dissatisfaction and disloyalty, and negative attitudes towards the company or product. This study focuses on SerQ as a key component of overall system quality, incorporating dimensions like, reliability, empathy, guarantees and responsiveness.

(Harriet et al., 2024) highlighted the crucial role of empathy and reliability in determining service quality, particularly in terms of user retention and the consistent delivery of high-quality services. These findings further emphasize the importance of empathy and reliability as key components of SerQ, which can significantly impact user satisfaction and organizational success. Table *I* summarizes the literature review and shows the key finding for each approach.

Table I
Summary of Literature Review

| Reference | Key Findings |
|---|---|
| (Knauer et al., 2020) | SQ is an important factor when it comes to information system success. |
| (Oktrivina et al., 2021b) | SQ includes system reliability, response time, and usability. |
| (Ferreira et al., 2020) | Usability and functionality influence user satisfaction and perceive usefulness. |
| (Hasan et al., 2014) | Reliability and performance are two sub dimensions of system quality, which are critical to organizational performance. |
| (Dambanemuya & Diakopoulos, 2021) | IQ means accuracy, completeness, relevance, and suitability. |
| (Serrano-Aguilar et al., 2021) | Information flow is an essential element of decision-making and organizational performance since it must be accurate and timely. |
| (Kalankesh et al., 2020b) | System efficiency and user satisfaction is influenced by relevance information and positively comprehensiveness. |
| (Ighomereho & Omoyele, 2022) | SerQ includes: Responsiveness, assurance, and empathy. |
| (Kevin Fuchs, 2021) | Responsiveness and assurance are vital dimensions in assessing service quality. |
| (Zhou et al., 2020) | Empathy and reliability are crucial factors in determining service quality, particularly for user retention. |

### III. RESEARCH METHODOLOGY

#### A. Research Design:

A quantitative study design was employed to explore the interrelationships between SQ, IQ, and SerQ and their impact on organizational performance, utilizing a survey-based method to collect the primary data gathered via several samples of respondents across various industries. A structured questionnaire was developed to measure the three quality dimensions, incorporating validated scales from the literature. The questionnaire was administered electronically to a diverse sample of respondents, including IT professionals, business analysts, and end-users.

B. *Sample Selection:*

The sample for this study includes participants form diverse occupational backgrounds and educational levels, ensuring a comprehensive representation for perspectives relevant to quality assessment. The sample distribution is as follows:

- **Gender Distribution**:
    - Males: 62.5% (2236 participants)
    - Females: 37.5% (133 participants)
- **Age Distribution**:
    - 15% of them was in between 21-30 aged (60 participants)
    - 45% of them was in between 31-40 aged (180 participants)
    - 27.5% of them was in between 41-51 aged (110 participants)
    - 12.5% of them was in between 51-60 aged (50 participants)

This demographic diversity provides a robust basis for evaluating the quality metrics across different segments of the population.

C. *Data Collection:*

Three quality dimensions: SQ, IQ and SerQ have been selected to collect data developed by a structured survey. The research questionnaire had 23 questions that was measured on a Likert scale type. All the items were developed from existing scales in the literature to ensure reliability and measure validity. So as to calculate internal consistency of instrument measurements, Cronbach Alpha was determined and discovered to be (0.953), displays a comprehensive consistency of internals among other items which denotes that using the questionnaire is valid in measure the constructs.

D. *Data Analysis:*

By using SPSS, the collected data was analyzed, and R language utilized to perform various statistical tests including:

Descriptive Statistics: Taking advantages of some descriptive statistics like standard deviation, mean and distributing frequency to discover the information demographics of the respondents and the responses to the survey questions.

Reliability Analysis: Using "Cronbach's Alpha" in order to evaluate the internal (local) measurement scale reliabilities, recording the result of 0.953 presents that the system is in a high level in terms of reliability.

To determine the properness of the data in terms of "factor analysis" two tools were applied, "Kaiser-Meyer-Olkin (KMO)" to measure and "Bartlett's Test of Sphericity".  As, KMO value which resulted 0.965 and Bartlett's Test (p<0.000), this is show that the current research was determined to be high, this is also shows the suitability of the data for Anderson's test.

*Statistical Tests:*

Relay on the research questions to test the study hypotheses there are several statistical tests were employed presents bellow:

- Reliability Statistics: The measurement consistency of instrument was assessed utilizing Cronbach's Alpha. The 0.953 alpha result coefficient specifies a high level in terms of internal consistency among other items, suggesting that the study questionnaire is reliable measure of the constructs.
- Furthermore, the corrected item-total correlations (CIC) arranged from 0.55 to 0.74, shows that all items can significantly shares to the scale completely. Additionally, the Cronbach's Alpha remained consistently high (0.951-0.953) when items were removed individually, suggesting that no single item substantially impacts the total reliability of the scale.
- KMO and Bartlett's Test: KMO which measure sampling and Bartlett's Test of Sphericity was conducted to estimate the appropriateness of the data for making sure of factor analysis. (0.965) value of KMO represents the suitability of samples for factor analysis highly. And, Bartlett's Test of Sphericity was significant ($p < 001$) this is propose that the "correlation matrix" is not an identity matrix, therefore this is suitable for "factor analysis".
- Factor Analysis: Principal Component Analysis (PCA) with Varimax rotation was used to identity the highlighted dimensions of the constructs, two main components have been extracted through analysis phase, which explained in an important portion of the variance in the data. Component 1, accounting for 29.534% of the variance, while Component 2 accounts for 25.182% of the variance.

E. *Data Visualization:*

To visually represent the data and facilitate interpretation, Python libraries such as Matplotlib, Seaborn, and Pandas were used. Bar charts, donut charts, pie charts, and heatmaps were created to visualize reliability statistics, item-total correlations, and factor loadings. Communalities and variance explained tables were generated to provide visions into the diverse amount of each variable to be calculated, 216 for common factors. These tables help to identify the most effective items in capturing the constructs of SQ, IQ, and SerQ.

As indicated in Fig 1, the use of the scientific research method in this work enhances reliability and validity of the results. This method encompasses the use of well-structured questionnaires, use of sample diversity and good statistical analysis. The high Cronbach's Alpha coefficients also provide the validation of the measurement scale used in the research and the factor analysis also reveals a good fit providing further credibility to the data collected. The respondents from different industries and occupational levels make the study findings more generalizable. The study therefore offers an evaluation of SQ, IQ, and SerQ and in so doing, presents an understanding of the factors that affect information system performance and user satisfaction.

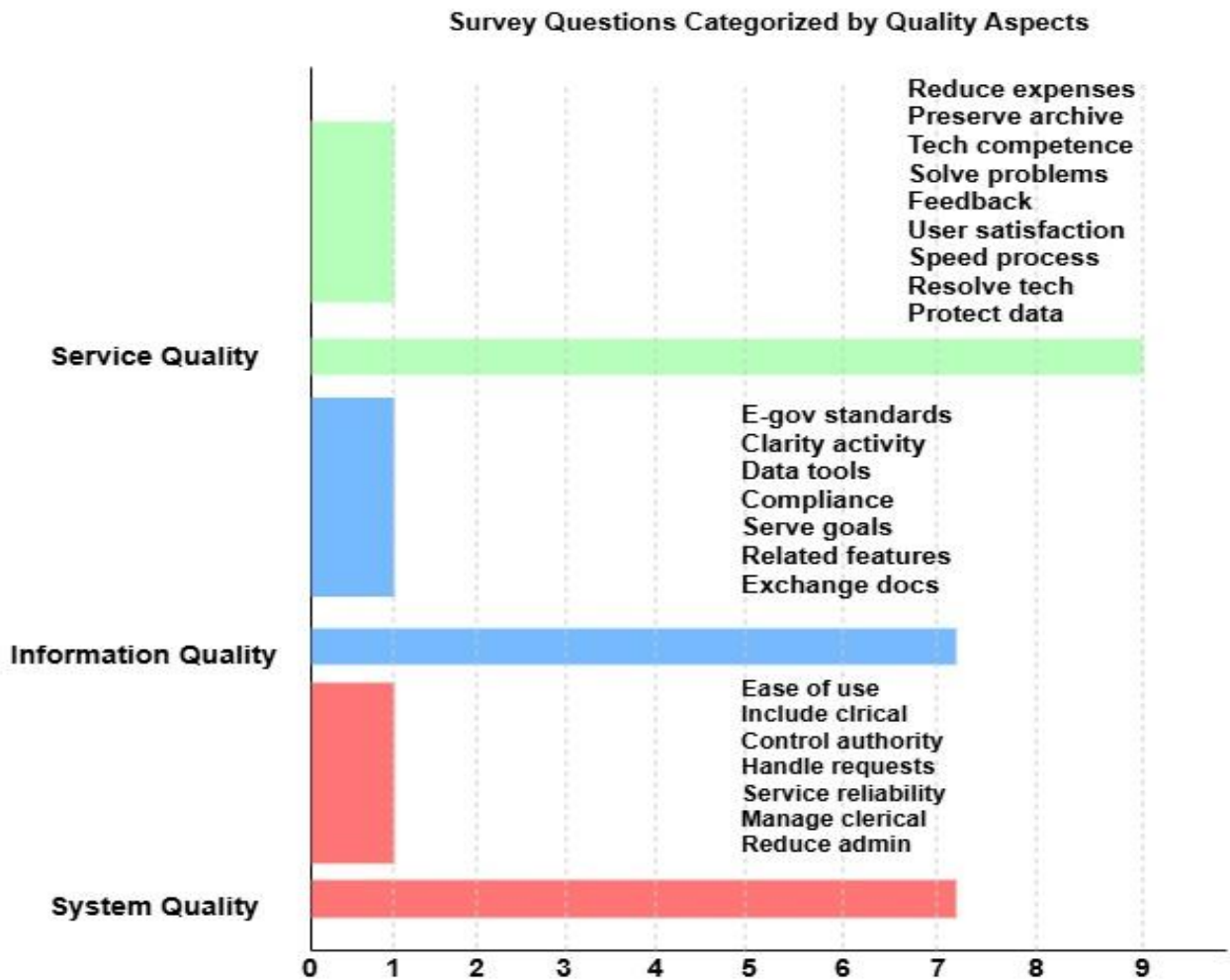

Fig 1 : Survey Quality Questionnaire

## IV. VISUALIZE STATICAL DATA

From a gender perspective, male participants dominated this study with (236) of them representing (64.0 %) of the sample while female participants were (133) representing (36.0 %) of the sample. This difference in gender may point a direction to sampling possibly meaning that the sample has a bias. With regards to the age distribution, 40.1% of participant fell within the age range of 26 – 35 years. This implies that young professionals were well represented in the study was a good cross-section representation of working young professionals. The middle career professionals in the 36-45 years age bracket comprised 29.9% of the sample.The remaining age groups, 18-25 and 46-55, were less represented, with 20.1% and 9.9% of the sample, respectively.

From Fig 2 below, this demographic profile may imply that the study results might be skewed towards young professional participants in the survey. This could in one way or the other reduce the generality of the study findings to other age professionals or other people of different ages. To fix this, future research needs to recruit a higher number of people who come from different age brackets and many different demographics.

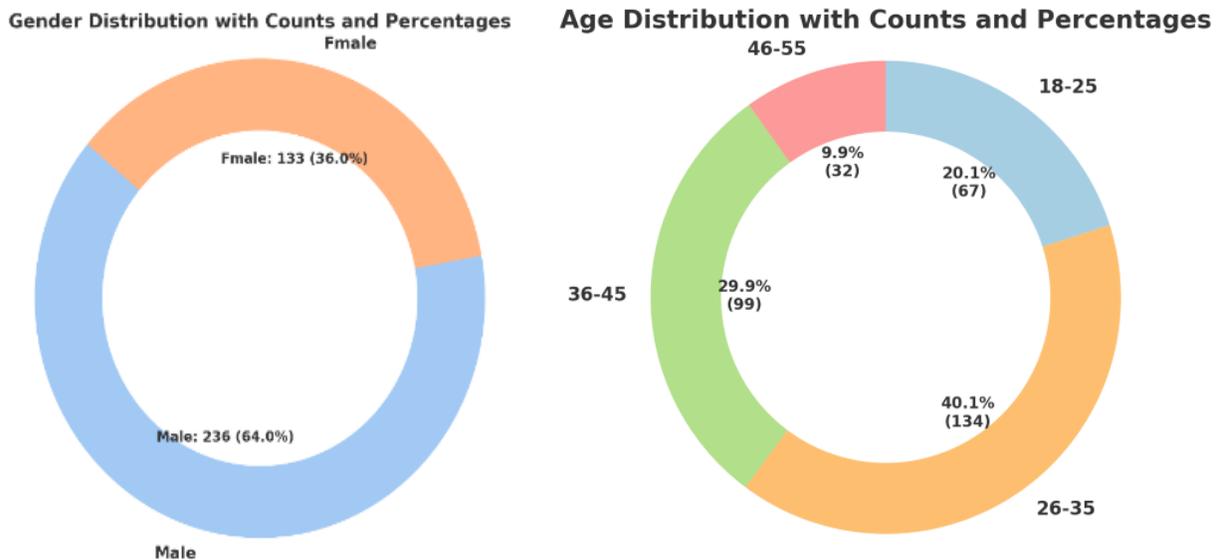

Fig 2: Demographic Profile of Respondents

In the current study, the occupation, specialty, and educational level of the participants are described in Fig 3 as follows. The sample includes the administration (68.3%), faculty (28.7%), and others (3.0%). It also guarantees that there is a wide range of approaches to the quality issue to make sure that all the aspects are considered.

With regard to specialization, the sample hardly includes those who specializes in a scientific field and a scientific-research activity (48.2%), humanities (43.1%), and the others (8.7%). This diversity of expertise enriches the data and provides the opportunity to describe quality factors more comprehensively.

The participants' educational level is ranked as follows: a bachelor's degree (64.2%) and a master's or doctoral degree (30.4%). While only 5.4% of the respondents have secondary education or express other qualifications. This high level of education implies that these respondents are in a good position to offer relevant feedback on information system quality and services.

All in all, the demographic characteristics of the sample suggest that a wide range of quality measures can be investigated without compromising the quality of the study. Indeed, the variety of professions, specializations, and levels of education contributing to the study guarantees the results' applicability to the general population.

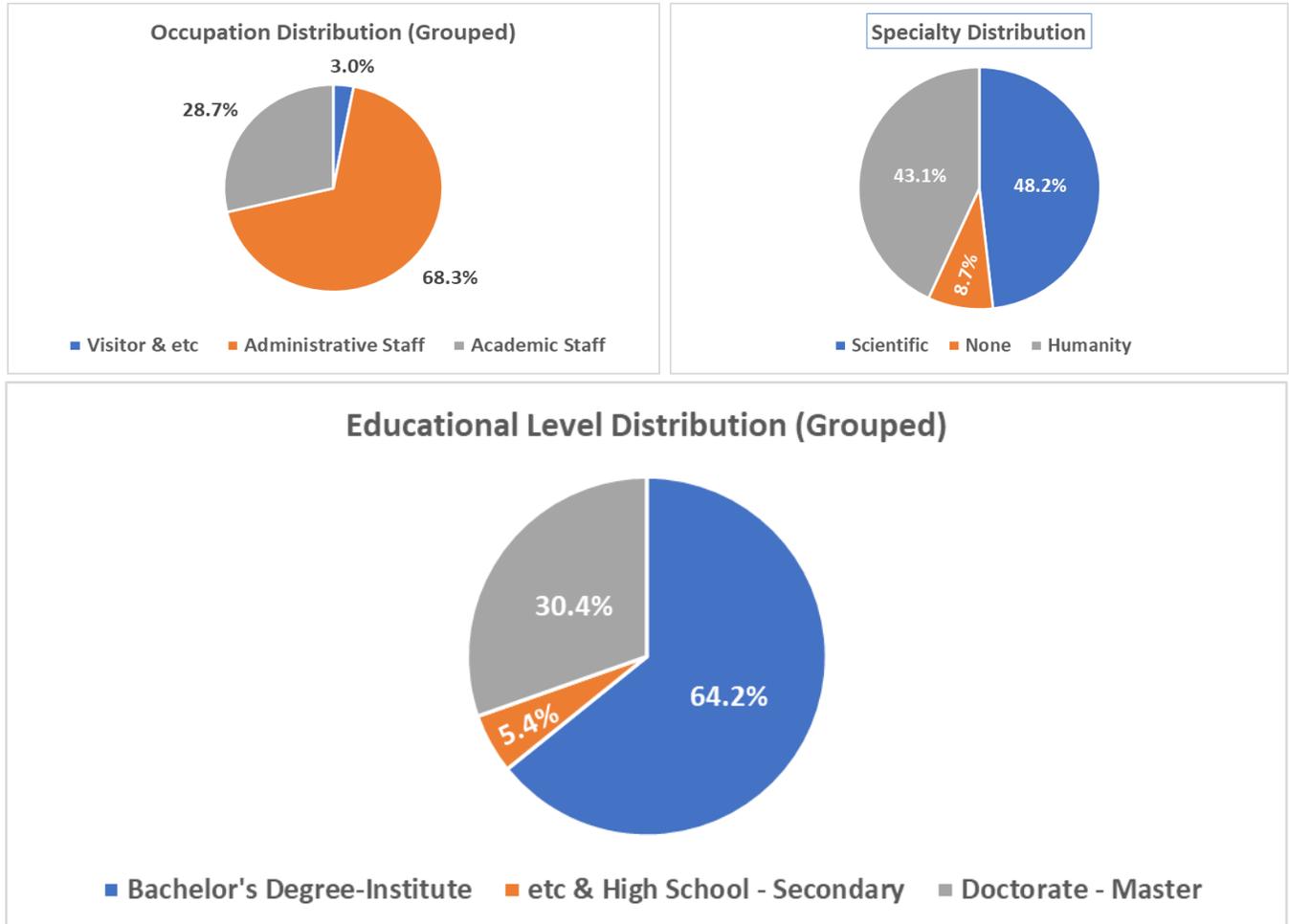

Fig 3: Distribution of Occupation, Specialty, and Education

The questionnaire development reliability statistics presented in Fig 4 bar has been utilized in the current work. On the internal or local reliability study, a "Cronbach's Alpha" coefficient of (0.953) for the 23-item measure was obtained giving an indication of a high reliability of the questionnaire to measure the constructs of SQ, IQ and SerQ. Taking the Cronbach's Alpha estimate a notch higher to (0.954) when standardized also strengthens the reliability of the scale. These high reliability coefficients make it possible for the study to have reliable result which increases confidence in the study findings.

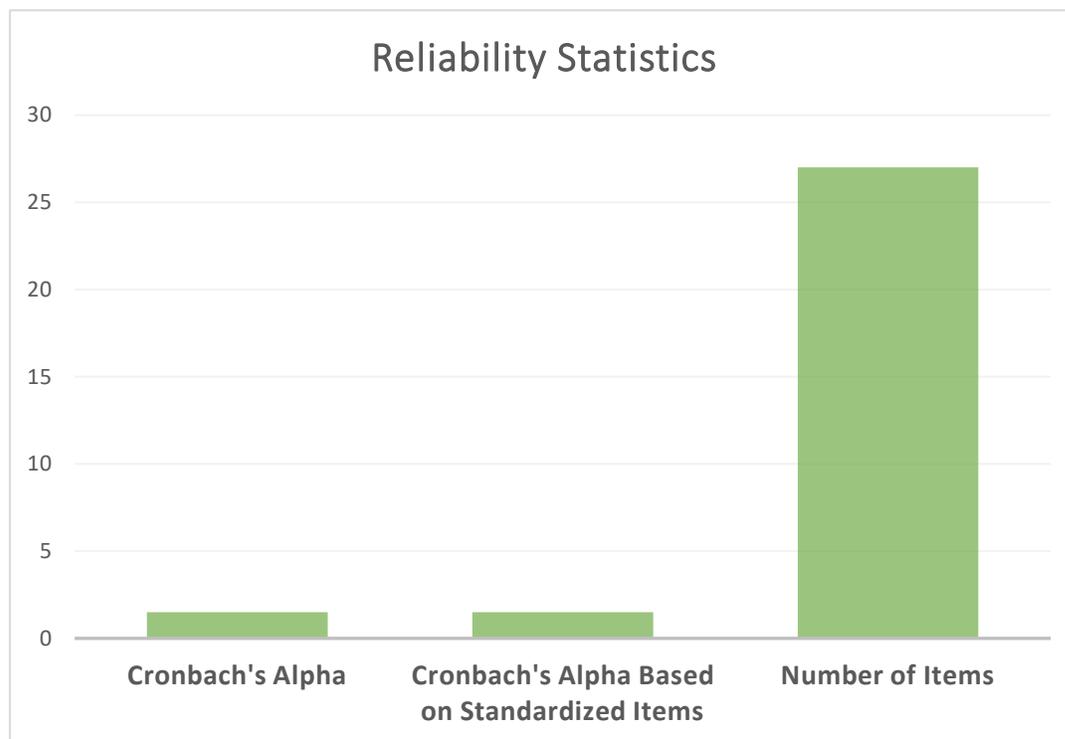

Fig 4: Reliability Statistics Explanation

Fig 5 Shows a heatmap of the CIC and "Cronbach's Alpha" when any item of the questionnaire has been deleted. And, the heatmap also represent the internal reliability and consistency of the use of instrument's measurement. The CIC calculate varies from 0.55 to 0.74, which suggests a good validity of the items for the overall construct in question. Higher values indicate that the item has a high degree of correlation with all scale scores. The Cronbach's Alpha if Item Deleted values are also above 0.9 (between 0.951 to 0.953) hence the delete model shows that no item reducing the internal consistency of the scale considerably. Thus, it can be suggested that the current scale is valid and reliable, despite the fact that some items were omitted. The heatmap successfully translates these relationships to assist in easily assessing the reliability of the measurement instrument.

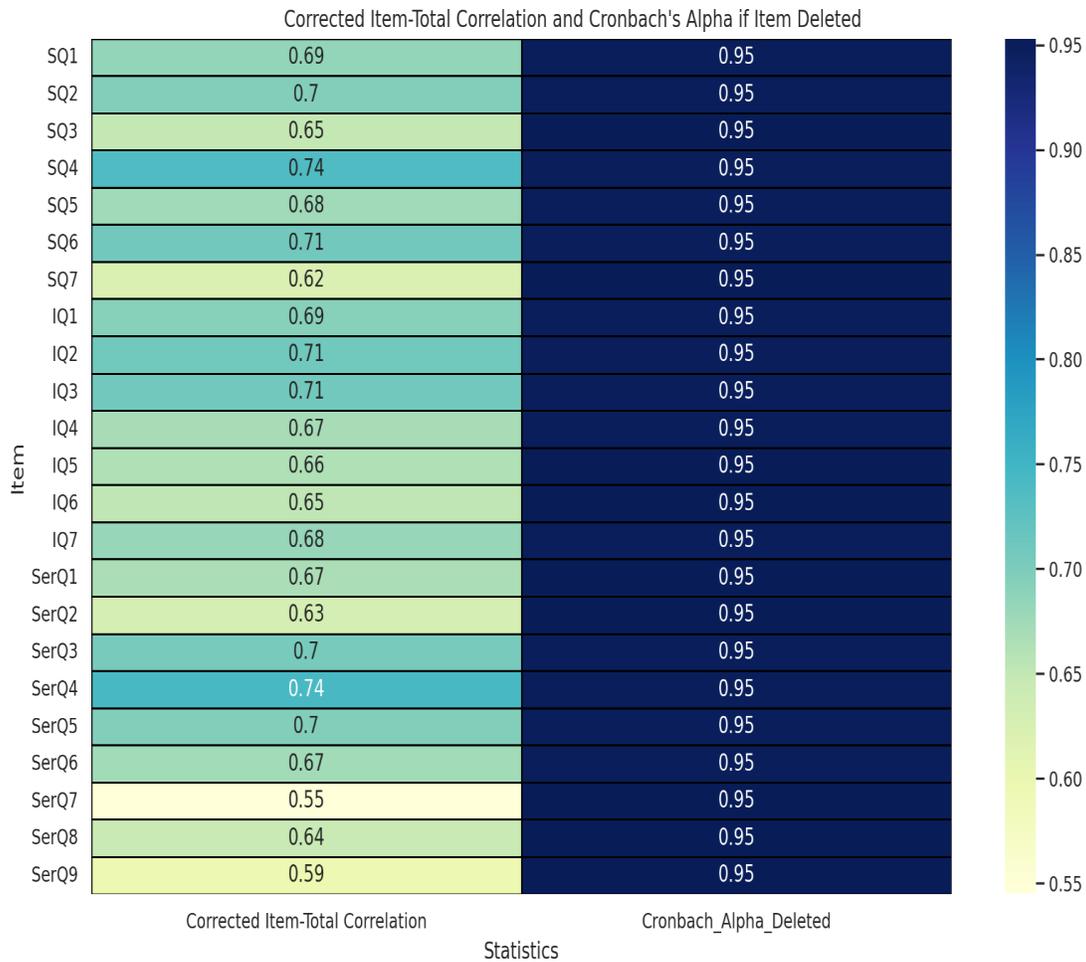

Fig 5: Explain Item-Total Correlation and Cronbach's Alpha.

**RESULTS:**

Table II
Reliability Statistics of Quality Aspects

| Cronbach's Alpha | Cronbach's Alpha Based on Standardized Items | N of Items |
|---|---|---|
| 0.953 | 0.954 | 23 |

As illustrated in *Table II*, the reliability assessment shows good internal consistency for the overall 23-item scale, through Cronbach's Alpha coefficient of 0.953. This value indicate that the items are positively affecting each other and indeed capture the constructs of SQ, IQ and SerQ. further, the exclusion of each item does not have any high impact on the Cronbach's Alpha co-efficient, which also declared that each item is relevant for contributing the reliability coefficient of the scale. These findings support the reliability and validity construct of the questionnaire.

Table III
Item's Contribution of Overall Reliability Scales

| | Scale Mean if Item Deleted | Scale Variance if Item Deleted | Corrected Item-Total Correlation | Squared Multiple Correlation | Cronbach's Alpha if Item Deleted |
|---|---|---|---|---|---|
| SQ1 | 85.9485 | 141.489 | 0.685 | 0.578 | 0.951 |
| SQ2 | 85.7724 | 141.377 | 0.696 | 0.570 | 0.951 |
| SQ3 | 85.9512 | 141.128 | 0.649 | 0.492 | 0.952 |
| SQ4 | 86.0569 | 139.668 | 0.737 | 0.593 | 0.951 |
| SQ5 | 86.0163 | 141.391 | 0.675 | 0.526 | 0.951 |
| SQ6 | 85.9404 | 141.192 | 0.709 | 0.563 | 0.951 |
| SQ7 | 85.7751 | 142.925 | 0.622 | 0.450 | 0.952 |
| IQ1 | 85.8130 | 140.506 | 0.692 | 0.573 | 0.951 |
| IQ2 | 86.0108 | 141.712 | 0.710 | 0.539 | 0.951 |
| IQ3 | 85.7344 | 140.603 | 0.708 | 0.594 | 0.951 |
| IQ4 | 86.0325 | 141.575 | 0.670 | 0.502 | 0.951 |
| IQ5 | 86.0650 | 141.615 | 0.663 | 0.494 | 0.952 |
| IQ6 | 86.0434 | 142.514 | 0.653 | 0.502 | 0.952 |
| IQ7 | 86.0136 | 140.975 | 0.682 | 0.542 | 0.951 |
| SerQ1 | 85.9160 | 141.104 | 0.667 | 0.510 | 0.952 |
| SerQ2 | 86.2276 | 139.671 | 0.631 | 0.481 | 0.952 |
| SerQ3 | 85.8238 | 139.640 | 0.703 | 0.615 | 0.951 |
| SerQ4 | 86.0569 | 139.532 | 0.744 | 0.604 | 0.951 |
| SerQ5 | 86.4092 | 138.704 | 0.696 | 0.612 | 0.951 |
| SerQ6 | 86.3821 | 139.405 | 0.674 | 0.582 | 0.951 |
| SerQ7 | 86.0136 | 143.883 | 0.545 | 0.393 | 0.953 |
| SerQ8 | 85.7127 | 141.901 | 0.644 | 0.481 | 0.952 |
| SerQ9 | 86.0786 | 141.328 | 0.595 | 0.437 | 0.952 |

*Table III* summarizes the reliability analysis of the items in the questionnaire so as to accurately assess the outcomes. The CIC coefficients are 0.55 to 0.74 for all the items which implies high correlation between each of the items and the overall scale. Additionally, the Cronbach's Alpha if Item Deleted values remain consistently high, suggesting that no single item significantly impacts the overall reliability of the scale. These findings are more likely to be locally supported or internally consistency and reliability of the survey.

Table IV
KMO and Barlett's Test

| **Items** | | **Item total range** |
|---|---|---|
| Kaiser-Meyer-Olkin Measure of Sampling Adequacy. | | 0.965 |
| Bartlett's Test of Sphericity | Approx. Chi-Square | 4927.038 |
| | df | 253 |
| | Sig. | 0.000 |

For the purpose of evaluating the appropriateness of the data for factor examination., the KMO measure of sampling adequacy and Bartlett's Test of Sphericity were conducted. As shown in the Table *IV* which is represent that the (0.965) as a KMO value indicate that the sample is very suitable for factor analysis.. Furthermore, Bartlett's Test of Sphericity was significant ($p < .001$), confirming that the correlation matrix is not an identity matrix and, therefore, factor scrutiny is appropriate.

Table V
Communalities Principal Component Analysis

|       | Initial | Extraction |
|-------|---------|------------|
| SQ1   | 1.000   | 0.617      |
| SQ2   | 1.000   | 0.607      |
| SQ3   | 1.000   | 0.470      |
| SQ4   | 1.000   | 0.586      |
| SQ5   | 1.000   | 0.560      |
| SQ6   | 1.000   | 0.581      |
| SQ7   | 1.000   | 0.435      |
| IQ1   | 1.000   | 0.585      |
| IQ2   | 1.000   | 0.550      |
| IQ3   | 1.000   | 0.588      |
| IQ4   | 1.000   | 0.534      |
| IQ5   | 1.000   | 0.485      |
| IQ6   | 1.000   | 0.508      |
| IQ7   | 1.000   | 0.555      |
| SerQ1 | 1.000   | 0.501      |
| SerQ2 | 1.000   | 0.525      |
| SerQ3 | 1.000   | 0.635      |
| SerQ4 | 1.000   | 0.599      |
| SerQ5 | 1.000   | 0.594      |
| SerQ6 | 1.000   | 0.649      |
| SerQ7 | 1.000   | 0.489      |
| SerQ8 | 1.000   | 0.507      |
| SerQ9 | 1.000   | 0.422      |

Table *V* presents the communalities for each item in the questionnaire. Communality represents the proportion of variance in each variable that is demonstrated by the common factors. Higher communalities indicate that the item is strongly related to the underlying factors.

As shown in the table, the initial communalities for all items are 1.000, indicating that all the change in each item is initially accounted for. After factor extraction, the extraction values range from 0.422 to 0.617, indicating that a substantial portion of the diversity in each item is explained by the common factors. These high extraction values suggest that the factor solution is adequate and that the items are well-represented by the underlying constructs.

Table VI
Variance Explained by Each Component

| Component | Rotation Sums of Squared Loadings | | | Total | % of Variance | Cumulative % |
|---|---|---|---|---|---|---|
| | Total | % of Variance | Cumulative % | | | |
| 1 | 11.467 | 49.855 | 49.855 | 6.793 | 29.534 | 29.534 |
| 2 | 1.118 | 4.861 | 54.716 | 5.792 | 25.182 | 54.716 |
| 3 | 0.960 | 4.176 | 58.892 | | | |
| 4 | 0.818 | 3.556 | 62.448 | | | |
| 5 | 0.772 | 3.357 | 65.806 | | | |
| 6 | 0.688 | 2.993 | 68.799 | | | |
| 7 | 0.639 | 2.780 | 71.579 | | | |
| 8 | 0.604 | 2.627 | 74.206 | | | |
| 9 | 0.578 | 2.515 | 76.721 | | | |
| 10 | 0.540 | 2.346 | 79.067 | | | |
| 11 | 0.519 | 2.256 | 81.322 | | | |
| 12 | 0.481 | 2.091 | 83.413 | | | |
| 13 | 0.461 | 2.002 | 85.416 | | | |
| 14 | 0.445 | 1.934 | 87.349 | | | |
| 15 | 0.415 | 1.802 | 89.152 | | | |
| 16 | 0.373 | 1.620 | 90.771 | | | |
| 17 | 0.359 | 1.561 | 92.332 | | | |
| 18 | 0.351 | 1.525 | 93.857 | | | |
| 19 | 0.329 | 1.431 | 95.289 | | | |
| 20 | 0.315 | 1.370 | 96.659 | | | |
| 21 | 0.269 | 1.168 | 97.827 | | | |
| 22 | 0.253 | 1.101 | 98.928 | | | |
| 23 | 0.247 | 1.072 | 100.000 | | | |

Table *VI* presents the consequences of the principal component analysis (PCA). The first two components explain a substantial portion of the variety in the data, with the first component accounting for 49.855% and the second component accounting for 4.861%. Together, these two components explain 54.716% of the total variance. After rotation, the first component clarifies 29.534% of the variance, and the second component explains 25.182%. This suggests that the variance is distributed across multiple factors, with the first two factors being the most important.

Overall, the PCA results indicate that the data can be effectively represented by a reduced number of underlying factors, suggesting a strong factor structure for the constructs of SQ, IQ, and SerQ.

Table VII
Rotation Method of Varimax with Kaiser Normalization

| | Component | |
|---|---|---|
| | 1 | 2 |
| SerQ3 | 0.749 | |
| SQ1 | 0.746 | |
| SQ2 | 0.721 | |
| IQ1 | 0.699 | |
| IQ3 | 0.679 | |
| IQ7 | 0.669 | |
| SQ6 | 0.666 | |
| SerQ8 | 0.645 | |
| SerQ9 | 0.572 | |
| SQ4 | 0.572 | |
| IQ2 | 0.538 | |
| IQ5 | 0.536 | |
| SQ7 | 0.472 | |
| SerQ6 | | 0.765 |
| SerQ5 | | 0.681 |
| SerQ7 | | 0.679 |
| SerQ2 | | 0.664 |
| SQ5 | | 0.658 |
| IQ4 | | 0.620 |
| IQ6 | | 0.602 |
| SerQ4 | | 0.550 |
| SerQ1 | | 0.549 |
| SQ3 | | 0.502 |

The rotated component matrix in Table *VII* offers insights into the underlying structure of the data. By rotating the factor axes, the interpretation of the factors becomes clearer. Component 1 is primarily related to SQ and IQ. Items such as SQ1, SQ2, IQ1, IQ3, and IQ7 have high loadings on this component, suggesting that they are strongly associated with these dimensions. Component 2 is more closely related to SerQ. Items like SerQ6, SerQ5, SerQ7, SerQ2, and SQ5 have high loadings on this component, indicating that they are strongly linked to service-related aspects. The clear separation of items across the two components supports the validity of the factor structure and suggests that the three quality dimensions (SQ, IQ, and SerQ) are distinct but interrelated constructs.

## V. OVERALL DISCUSSION

The analysis of internal consistency using Cronbach's Alpha reveals that all three quality dimensions (SQ, IQ, and SerQ) demonstrate high levels of reliability. The Cronbach's Alpha values consistently hover around 0.95, indicating that the items within each dimension are strongly correlated and measure a coherent construct. While all three dimensions exhibit strong internal consistency, it is important to note that the relative significant of each dimension may vary depending on the exact context and organizational goals. For example, in certain industries, System Quality may be a critical factor, while in others, Service Quality may be more important.

The communality values for the items within each quality dimension provide insights into the extent to which the features extracted from the analysis account for the variance in each item.

**SQ items**: The extraction values reached from (0.435) to (0.617), which means that quite a significant amount of the variability in these items results from the factors being tested. This indicates that the factors indeed account for the crucial aspects of System Quality.

**IQ items**: The extraction values vary between (0.485 and 0.588) which show that factors make a moderate to a strong contribution towards the explanation of the diversity in these items. This means that the factors sufficiently captured the main dimensions of Information Quality

**SerQ items**: The extraction values oscillate between (0,422 and 0,765) which indicates some variation of the factors in the items' variance. It was also observed that while extracting the factors the values of SerQ3 and SerQ6 were high, therefore indicating strong correlation with the factors, there are some items whose extraction values were less, implying that there may be need to extract some other few factors to give a full explanation of the variance.

All in all, it can be seen from the communality values that the factor analysis has indeed captured the essence of the three quality constructs.

The rotated component matrix also gives additional information in the structural aspect of the data. More specifically, Component 1 is nearly synonymous with what has been defined, previously, as SQ and IQ. Higher loadings on this component are SQ1, SQ2, IQ1, IQ3 and IQ7, which show a positive correlation with these elements. Component 2 seems to be more correlated with SerQ. On this component, high loading items include SerQ6, SerQ5, SerQ7, SerQ2, and SQ5, indicating that the aforementioned items are highly associated with service factors. In separating the items, the perfected ordinate and abscissa values demonstrate the factor structure and imply that the three forms of quality, namely, SQ, IQ, and SerQ are unique but correlated measures.

The analysis of the data reveals that the measures for (SQ), (IQ), and (SerQ) have high levels of reliability and validity.

**SQ:** The high internal consistencies along with the high factor loadings indicate that the components used in the SQ covers the technical features of the system, including functional, reliable, and usable characteristics.

**IQ:** This is evidenced by a high internal reliability and factor loadings of the IQ items in relation to the quality of the information obtained by the system in terms of accuracy, completeness, relevancy and timeliness.

**SerQ:** Furthermore, items of the SerQ also reveal acceptable values of internal consistency as well as of factor loading that indicates that this scale adequately captures the quality of service that is provided to the users. These constituents include; Responsiveness, assurance, and empathy.

In total, the results demonstrate the consistency and validity of the measurement tools employed in the analysis. These results have significant implications for research that seeks to explain what contributes to the performance and users' satisfaction with information system.

Therefore, based on the findings for all three quality dimensions of the research study, it can be concluded that both reliabilities, as well as validity is highly significant for SQ, IQ, as well as SerQ. Although SerQ has a slight advantage in a range of Q dimensions it may focus on.

Further, the high factor loadings for SerQ items within multiple components point to the fact that SerQ captures a broader class of service-related attributes, which range from responsiveness, assurance, empathy, to technical support. Thus, the coverage assures a more enhanced understanding of the quality of the services offered by the information system.

Although SQ and IQ also have good internal consistency reliability coefficients and factor loading, their focal areas could have been rather limited in scope. Thus, the analysis of the requirements matches the concept of SQ in which SQ is concentrated on the technical aspect of the system like functionality, reliability and usability. IQ, on the other hand focuses on the correctness, comprehensiveness and the related and needed information produced.

In light of these findings, this research has the following significant implications for organizations wishing to enhance the quality of their information systems. In this way, by implementing all of the mentioned conceptual quality dimensions, an organization is able to improve its user satisfaction, system performance, as well as organizational performance. Specifically, organizations should:

- **Focus on SerQ**: To ensure a positive user experience, attention should be given to aspects such as technical support, problem-solving and responsiveness.
- **Maintain SQ:** Ensure that the system is reliable, efficient, and user-friendly.
- **Prioritize IQ**: Focus on the precision, completeness, and relevance of the information provided by the system.

## VI. CONCLUSION

Hence, the need to combine SQ, IQ, and SerQ as main criteria for assessing the performance as well as reliability of information systems is stressed in this research. This study is therefore useful in establishing the interaction between these quality metrics toward benefiting practitioners and researchers. The studies present evidence that the three dimensions of quality are important for the success of information systems. There are considerable differences between the dimensions but they all affect each other, in some way. For instance, high quality system (SQ) may support production of accurate and timely information (IQ), which may improve quality-of-service delivery (SerQ).

The quantitative research used in this paper was supported by the statistical analysis method, which offered conclusive and empirical results. Thus, high consistency and validity of the measurement instruments guarantee the correctness of the outcomes. To further enhance the quality of information systems, organizations should consider the following strategies:

- **Prioritize Service Quality**: To improve the skills and knowledge of service providers the investigation in programs have to be done.
- **Ensure Information Quality**: For maintaining consistency, data accuracy and the completeness the implementing of data governance practices has to employed.
- **Optimize System Quality**: For ensuring the reliability and optimal performance the updating and maintaining systems has to be done regularly.

Thus, the three mentioned concentrations will help focus the efforts of organizations on providing enhanced satisfaction of users, the optimization of organizational operations, and increased organizational performance.

Further works should embrace how the quality improvement affects the organizational measures in the future more specifically, the financial aspect and the number and loyalty of the customers. Furthermore, research concerning the impact of technologies in improving information system quality including the artificial intelligence and machine learning would be another worthwhile interest. To answer these research questions, future research will be useful in helping to advance the continued development of high-quality information systems.